 \documentclass[final,authoryear,3p,times,twocolumn]{elsarticle}

\usepackage[T1]{fontenc}
\usepackage{multicol}
\usepackage{setspace}
\usepackage{tocloft}
\usepackage{amsmath,bm}
\usepackage{listings}
\usepackage{caption}
\usepackage{graphicx}

\usepackage{verbatim}

\usepackage{multirow}
\usepackage{longtable}
\usepackage{float} 
\usepackage{booktabs}

\usepackage{fancyhdr}

\usepackage{amssymb}
\usepackage{float}
\restylefloat{figure}

\journal{}

\begin{document}

\begin{frontmatter}



\title{Coupled Oscillators as a Model of Olfactory Network. Importance in Pattern Recognition and Classification Tasks}


\author{Alexandra Pinto Castellanos}
\address{Zurich, Switzerland}
\address{apintoca@student.ethz.ch}
\date{September 2017}

\begin{abstract}
The olfactory system is constantly solving pattern-recognition problems by the creation of a large space to codify odour representations and optimizing their distribution within it. A model of the Olfactory Bulb was developed by Z. Li and J. J. Hopfield \cite{Li1989} based on anatomy and electrophysiology. They used nonlinear simulations observing that the collective behavior produce an oscillatory frequency. Here, we show that the Subthreshold hopf bifurcation is a good candidate for modeling the bulb and the Subthreshold subcritical hopf bifurcation is a good candidate for modeling the olfactory cortex. Network topology analysis of the subcritical regime is presented as a proof of the importance of synapse plasticity for memory  functions in the olfactory cortex.
\end{abstract}

\end{frontmatter}


\section{Introduction}
The olfactory system is constantly solving pattern-recognition problems by the creation of a large space to codify odour representations and optimizing their distribution within it. Odours are multidimensional objects, mostly found as mixtures of molecules \citep{Laurent2002}. They are noisy and variable however we are able to classify them pretty well given the olfactory mechanism where two complementary process are present:
\begin{itemize}
\item Slow non-periodic process for decorrelation since most receptor cells respond to more than one odor type .
\item Fast oscillatory process given that the subcomponents of the mixture will bind to different receptor, the information about any odor maybe distributed across all receptors\citep{Li1990}.
\end{itemize}
After this, the olfactory centers integrate the information in order to recognize individual odors. On the other hand, it is able to identify a new odor in a mixture and memorize it for future recognition. 
\subsection{Biological Network}
The Olfactory Bulb is composed of the bulb and the pyriform (primary olfactory cortex) which send and receive feedforward and feedback information between each other. The bulb has excitatory activity created by mitral cells which receive input from olfactory receptors and generate oscillatory pattern and inhibitory activity  created by granule cells which modulate activity pattern of mitral cells. The collective behavior of this interconnection has a coherent oscillatory frequency of 35-90 Hz, however bulb's input is not oscillator\citep{Lancet1986}. The bulb sends its oscillatory output information to the pyriform. The pyriform has excitatory activity created by pyramidal cells and inhibitory activity created by  feedback inhibitory activity created by interneuron cells that oscillate at similar frequencies to the bulb. The pattern of activity are decision states of the odor information made by the bulb\citep{Li1990}.
\subsection{Artificial Network}
The olfactory bulb has three special properties that are appealing for implementations in pattern recognition. First, the fact that the same neural time series is recovered whenever a stimulus of a learned input is presented. Second, it is the most accessible and well preserved sensory system across vertebrates. Third, it's dynamics creates multiple stable states that were formed as a sort of associative memory and that can be used as templates for future recognition in the presence of input patterns\citep{Laurent2002}.\\ \\
The connections between mitral and granule cells are modeled by a coupled oscillator system. Where the eigenvectors and eigenvalues of the coupling matrix A determine the solutions of the oscillation mode, amplitude and phase. The whole bulb will oscillate with the same frequency given that the fastest growing mode will dominate the output. The coupling matrix will determine the oscillation and only if it is asymmetric, the oscillations can arise. The coupling matrix is variable depending on the input, causing an intermittent oscillating pattern with outputs of slow (2-4 Hz) and high (25-60 Hz) frequency component\citep{Yao1990}.\\ \\
On the other hand, the pyriform is modeled by a single dissipative oscillator driven by an oscillatory force. This oscillatory stimulus will resonate with the pyriform if its natural frequency is the same to the one of the oscillator.  \citep{Yao1990}.
\section{Related Work}
The first person who proposed an oscillating system as a  pattern recognition tool for the olfactory bulb was Bill Baird\citep{Baird1986} he proposed the use of single or double Hopf bifurcation in one or two oscillators to make stable cycles occur and the simulation of such cycles was required for modelling the olfactory bulb. A problem with this model was that, he had to force the system to produce them by using excitatory excitatory connections in the mitral cells, however, this connections are hard to find in the olfactory bulb. He used multiple Hopf bifurcations to create a system that makes the closed orbits possible and with them the selection of specific classes of inputs. \\ \\

Afterwards, a model of the Olfactory Bulb was developed by Z. Li and J. J. Hopfield \cite{Li1989} based on anatomy and electrophysiology. They used nonlinear simulations observing that the collective behavior produce an oscillatory frequency of 35-60Hz across the bulb. Almost parallel to this research, Yong Yao and Walter J. Freeman \cite{Yao1990} developed their research in the dynamics of the olfactory system as coupled nonlinear differential equations in order to understand the role of chaos for its pattern recognition function. Here, they found that the system maintains a low dimensional global chaotic attractor that create a stable state  ready to be accessed for pattern recognition.

\section{Model}
The model requires the following components;
\subsection{Coupled nonlinear equations. Oscillatory behavior}
A model of the olfactory bulb is implemented in a network of hopf systems that contains both excitatory and inhibitory node activity distributed equally in two populations of coupled excitatory and inhibitory nodes, one for modeling the bulb and the other for modeling the pyriform. The nodes in each population are composed of pairs of excitatory and inhibitory activity, corresponding to mitral and granule cells, where they are interconnected in a negative feedback loop. \\\\
Each node independently, this means without coupling, behaves as damped local oscillator. Once the coupling condition holds, the oscillation patterns will have specific amplitudes for each oscillator and specific phase relations between them, but the same frequency, they will have resonant collective behavior. Then, using phase plane methods and numerical solutions it is possible to obtain conclusions with respect to the effect of the input on the oscillating network. 
\subsection{Hopf-Bifurcation}
The understanding and usage of hopf bifurcations is going to be crucial in modelling the olfactory bulb, given the necessity for close orbits. Lets start the discussion of a Hopf bifurcation by defining that a bifurcation occurs when a stable equilibrium or closed orbit change their stability by when a parameter is change. A bifurcation is defined in terms of the eigenvalues of our system as the process when the real part of the eigenvalues is less than zero but then cross into the positive real plane \citep{Strogatz2011}.\\ \\

There are two types of bifurcations in fixed points depending on the eigenvalue: $\lambda=0$ and $\lambda \pm i\omega$, the later case is our case of interest because is the hopf bifurcation and this lets to the creation of close orbits. This orbits are going to be important in pattern recognition and classification tasks. Is this closed orbit what gives the stability to the system, otherwise the oscillation would be unstable given that the nature of the oscillator is damping if uncouple.\\ \\   

The system that is presented in this report is the case of two subcritical hopf coupled systems and latter on the coupling of four with different connectivity. First we start by analyzing the equation of the system, and how it behaves, this means the behavior of a Hopf-type system when it is coupled. Let's analyze the equation of the system, and how it behaves:
\begin{equation}
\dot{z}=(\mu+i)\omega z-\omega |z|^2z
\end{equation}
This equation is better visualized if we separate its real and imaginary parts. Assuming that $z=x+\rm{i}y$, we get:
\begin{align}
\dot{z}&=\dot{x}+i\dot{y} \\
\dot{x}&=-x^3\omega -xy^2 \omega +x\mu\omega -y \omega \\
\dot{y}&=-y^3\omega -yx^2 \omega + y\mu\omega +x \omega 
\end{align}

\begin{figure}[h]
\centering
\includegraphics[width=0.4\textwidth]{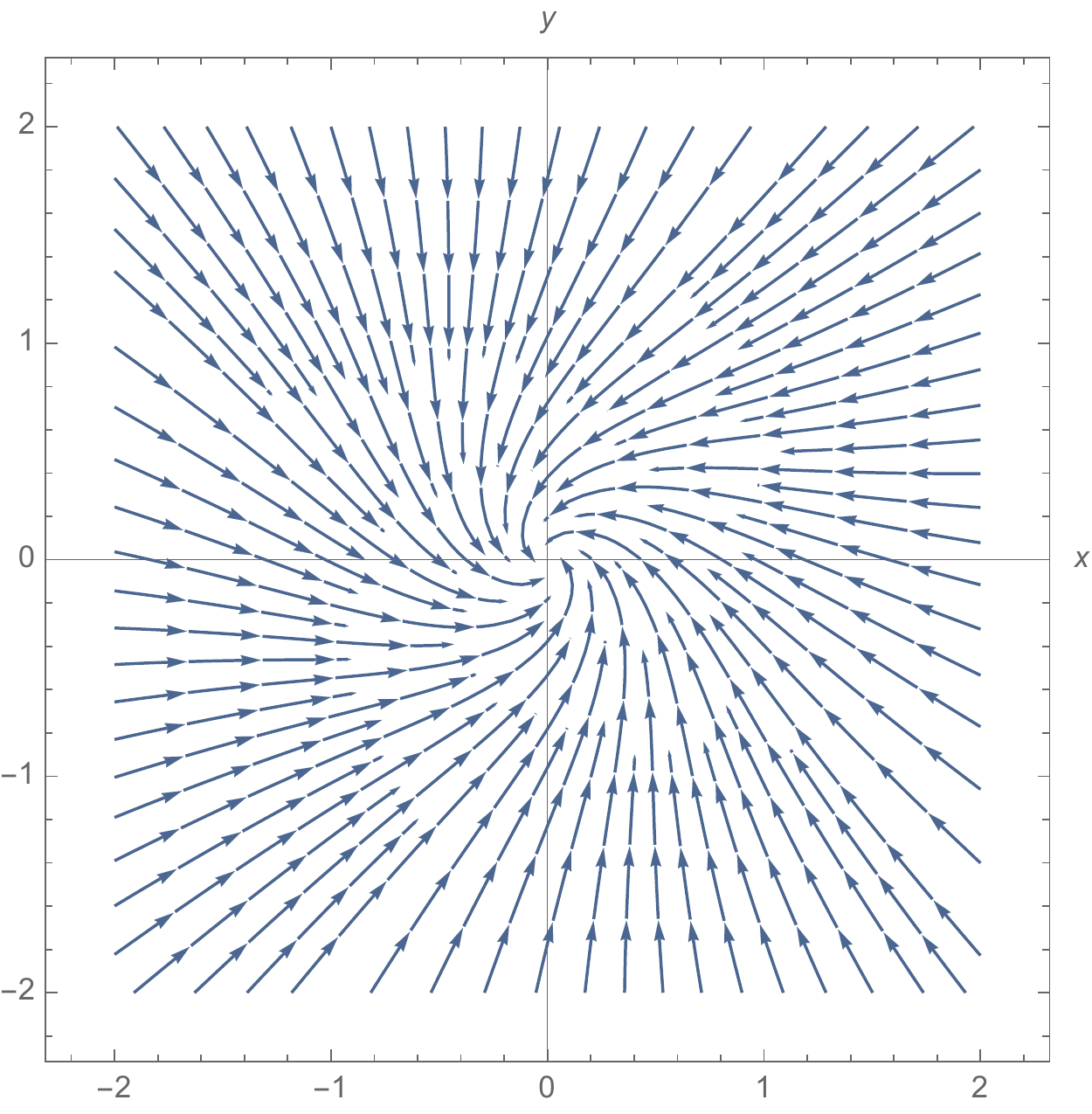}
  \caption{Behavior of the Hopf system sub threshold $\mu<0$ and $\omega=1$}
  \label{fig:sfig1}
\end{figure}

\begin{figure}[h]
\centering
\includegraphics[width=0.4\textwidth]{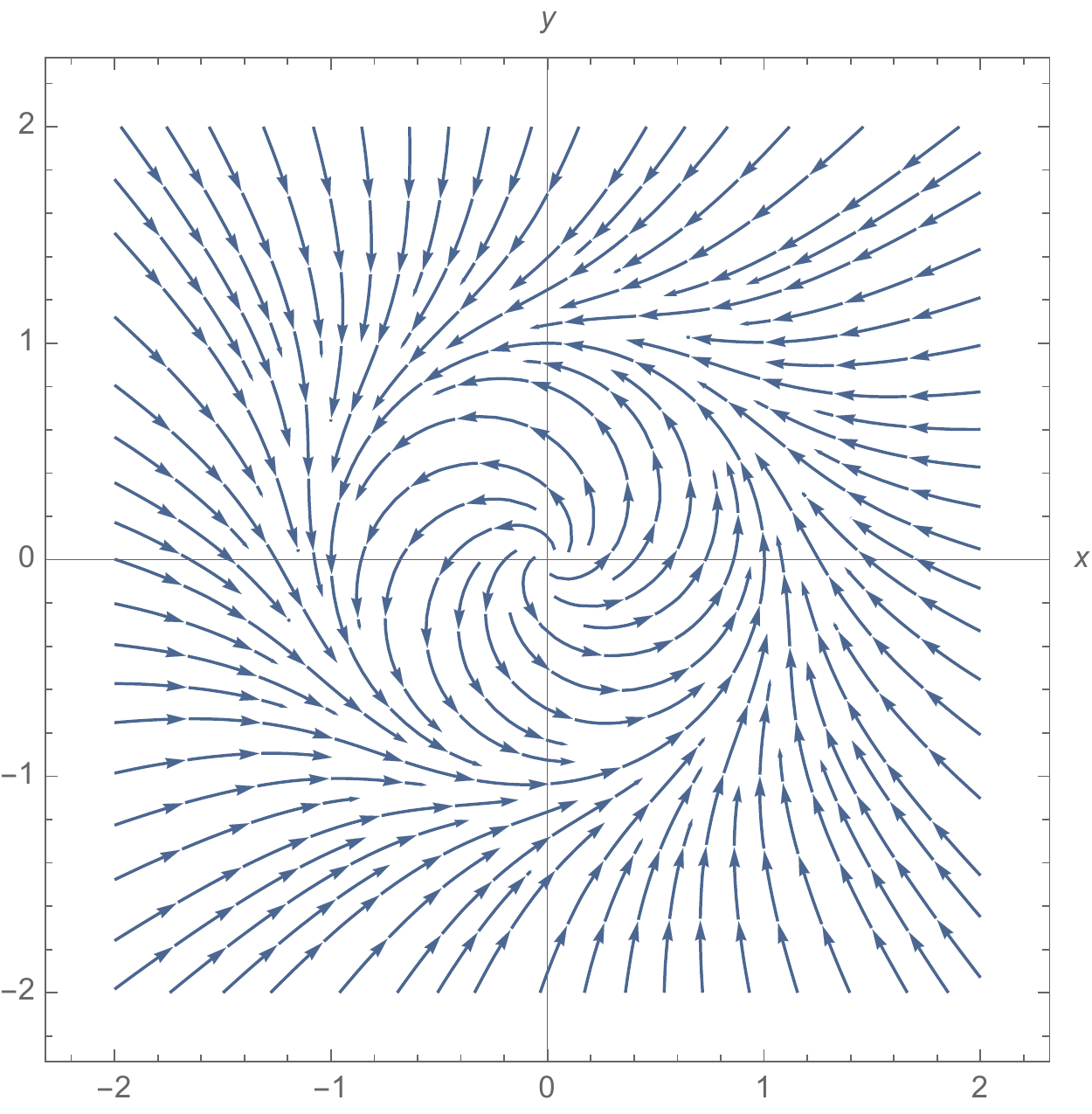}
  \caption{Behavior of the Hopf system above threshold $\mu>0$ and $\omega=1$}
  \label{fig:sfig2}
\end{figure}

\begin{figure}[h]
\centering
\includegraphics[width=0.4\textwidth]{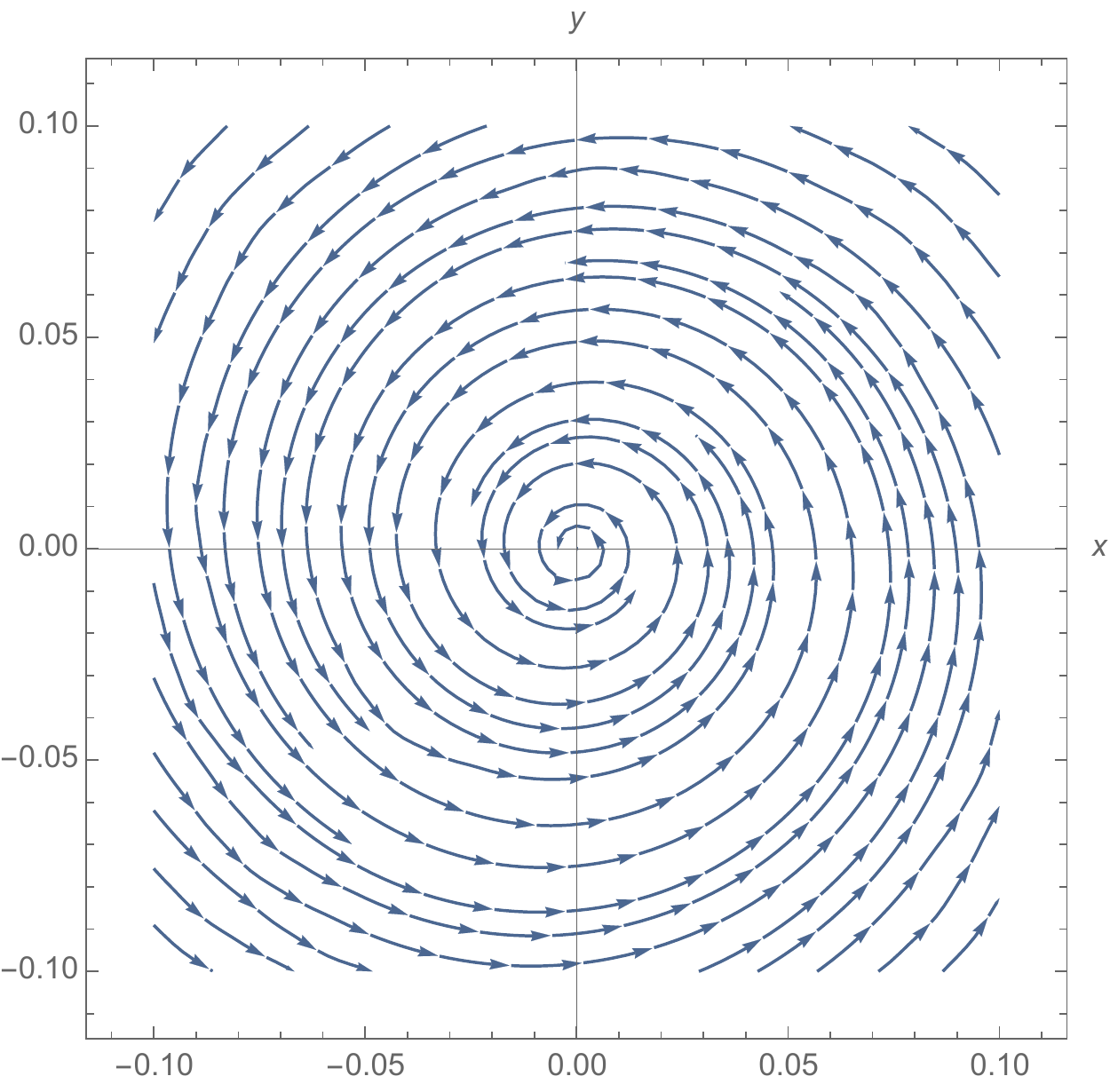}
  \caption{Hopf Systems $\mu=-0.1$ and $\omega=180\pi 2$}
  \label{fig:H180}
\end{figure}

\begin{figure}[h]
\centering
\includegraphics[width=0.4\textwidth]{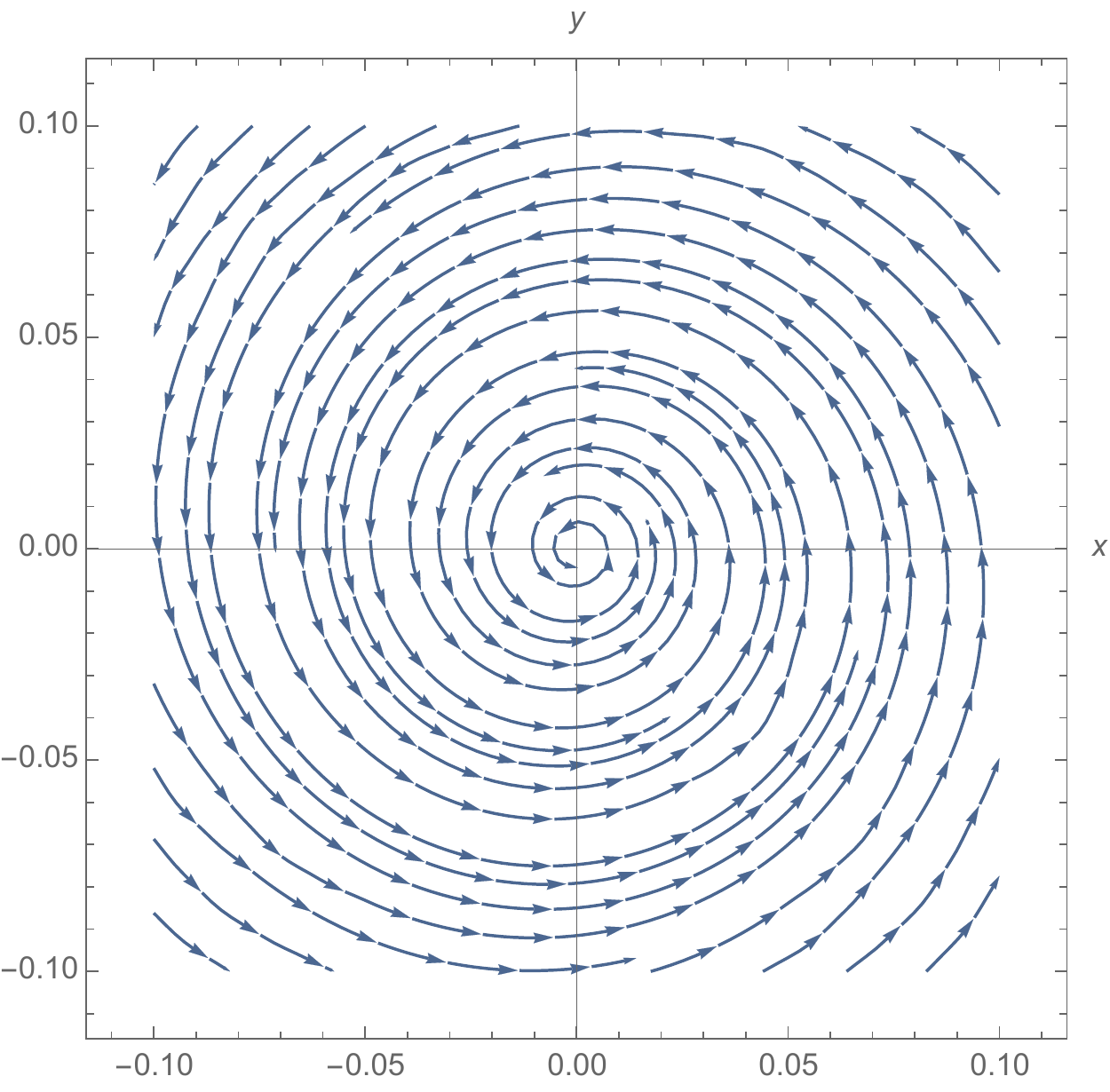}
  \caption{Hopf Systems $\mu=-0.1$ and $\omega=225 \pi 2$}
  \label{fig:H225}
\end{figure}

By studying $\dot{x}$ and $\dot{y}$. We can see that we have two possible states for this system, separated by a Hopf bifurcation at $\mu=0$. The sub-threshold state (fig. \ref{fig:sfig1}), in which the system decays regardless on the initial conditions, with $\mu<0$. And the above-threshold state (fig. \ref{fig:sfig2}), in which the real and imaginary part of the system oscillate in a stationary state, and the magnitude of $z$ stays constant. Additionally, for $\mu>0$ there is an unstable stationary point at $(x,y)=(0,0)$ and a circular limit cycle at $\sqrt{x^2+y^2}=r=\sqrt{\mu}$, in polar coordinates.
The system under interest are a sub-threshold Hopf systems, in which the input signal is a weak signal, this is the desirable case given that the amplitude of odors is really small, the system will decay to zero if the input is turned off. The system will have $\mu < 0$, $\omega_1=180\cdot 2 \pi$ and $\omega_2=225 \cdot 2\pi$, their behavior can be seen in figures \ref{fig:H180} and \ref{fig:H225}, both system are relatively similar given their proximity in frequency, and they decay to zero overtime to their fixed point if there is not an input signal as it's desirable.

\subsubsection{Sub-threshold Hopf System}
We should center our attention to the subthreshold case for modelling the behavior of the bulb, in the following section we will analyze the model for the pyriform. Now the coupling of these system, when a small input signal is fed:
\begin{align}
\dot{z}_1&=\omega_1\left[(\mu+i) z_1-|z_1|^2z_1+\frac{g_{21}}{2}z_2+h_0e^{i\omega t}\right]\\[1ex]
\dot{z}_2&=\omega_2\left[(\mu+i) z_2-|z_2|^2z_2+\frac{g_{12}}{2}z_1+h_0e^{i\omega t}\right]
\label{eq:sys2}
\end{align}
Here the coupling $g_{21}=g_{12}$, $h_0$ will have different small magnitudes and there will be a sweep in forcing frequency $\omega$, this oscillatory input is the representation of bulb's output and we have to check the response of the system under it's variation and given the coupling effect\\

The system is studied under different input amplitudes $h_0$, in this case it will be studied for 20dB, 40dB and 60dB. In the following figures the frequency response for different parameters can be observed. The upper curves in all plots are the 20dB response, the middle curves correspond to the 40dB, and the lower curves correspond to 60dB.\\

\begin{figure}[H]
\centering
\includegraphics[width=0.4\textwidth]{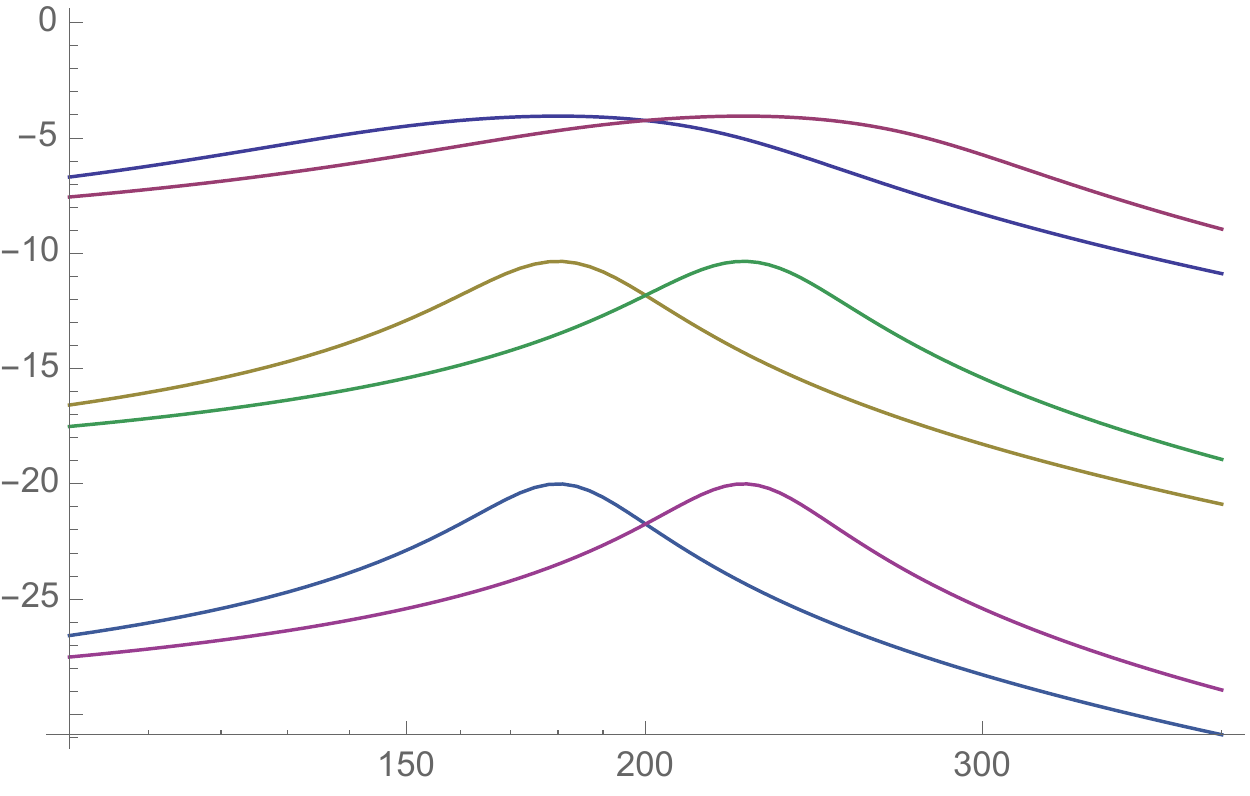}
  \caption{Behavior without coupling in Subthreshold $\mu=-0.1$ and $g=0$}
  \label{fig:C1}
\end{figure}

\begin{figure}[H]
\centering
\includegraphics[width=0.4\textwidth]{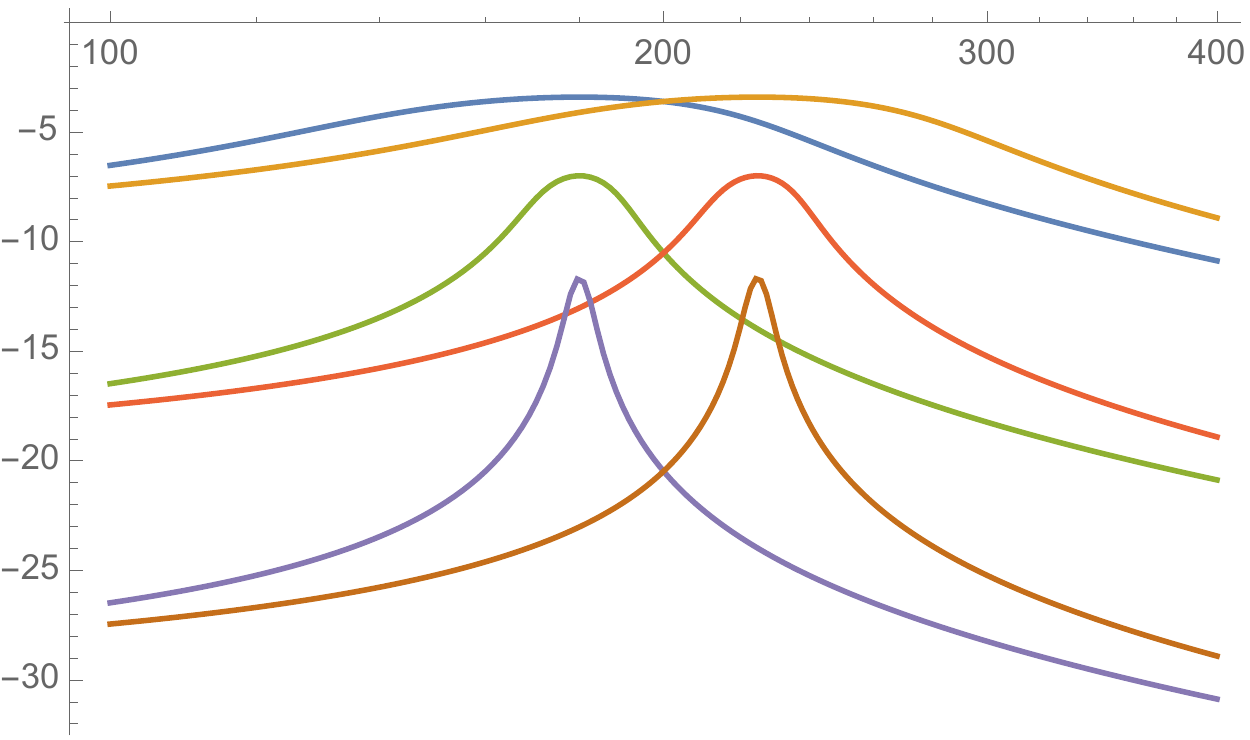}
  \caption{Behavior without coupling in Subthreshold $\mu=-0.01$ and $g=0$}
  \label{fig:C2}
\end{figure}

In figure \ref{fig:C1}, we can see the frequency response of the two oscillators with a coupling of zero and a damping term of $\mu=-0.1$, the same for figure \ref{fig:C2}, where the damping term is smaller. This reflects the importance of the coupling even in the presence of the input signal as in the case of the bulb where the signal goes to zero if there is no coupling. \\ \\

\begin{figure}[H]
\centering
\includegraphics[width=0.4\textwidth]{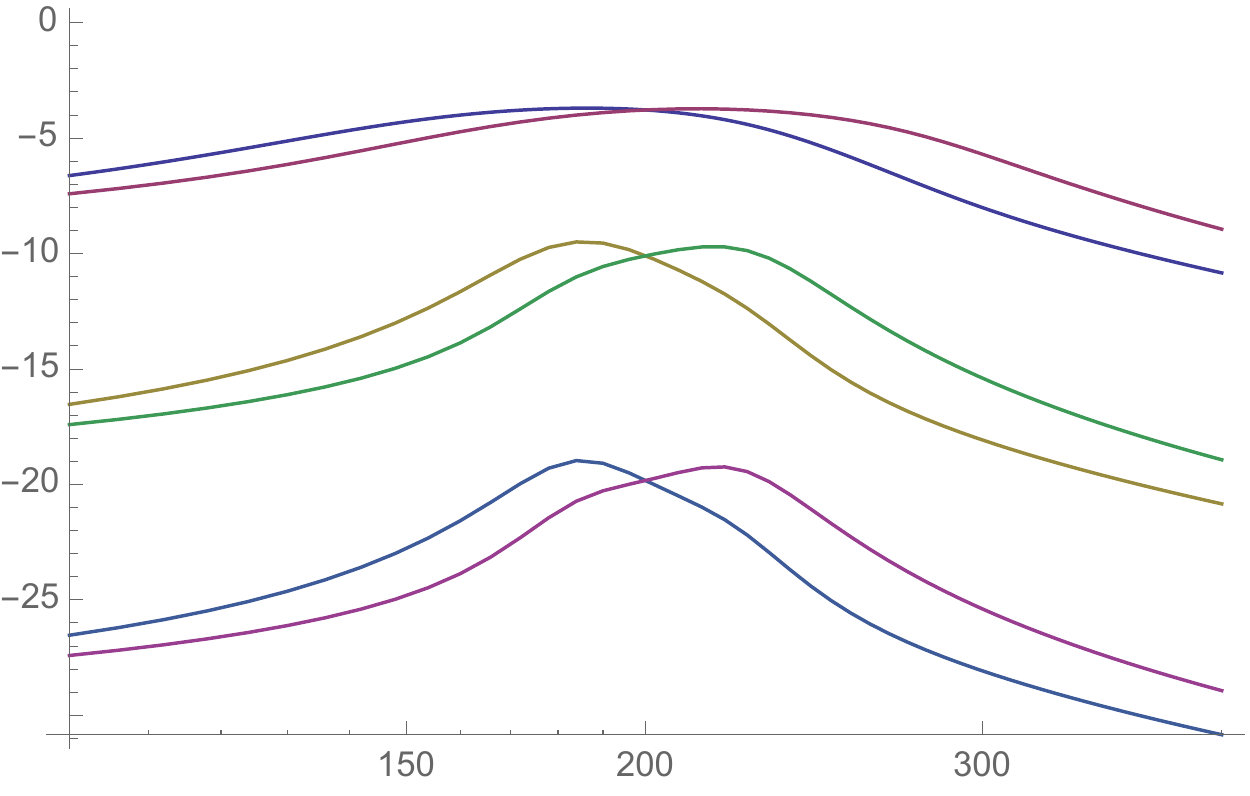}
  \caption{Influence of the coupling term in Subthreshold $\mu=-0.1$ and $g=0.12$}
  \label{fig:C3}
\end{figure}

In figure \ref{fig:C3}, we can see how the coupling term brings the oscillators closer in the resonant frequency. Is only in this case that the signal is transmitted to the olfactory cortex. However the coupling in figure \ref{fig:C3} is still not strong enough to cause a single resonant frequency of the two oscillators.\\ \\

\begin{figure}[H]
\centering
\includegraphics[width=0.4\textwidth]{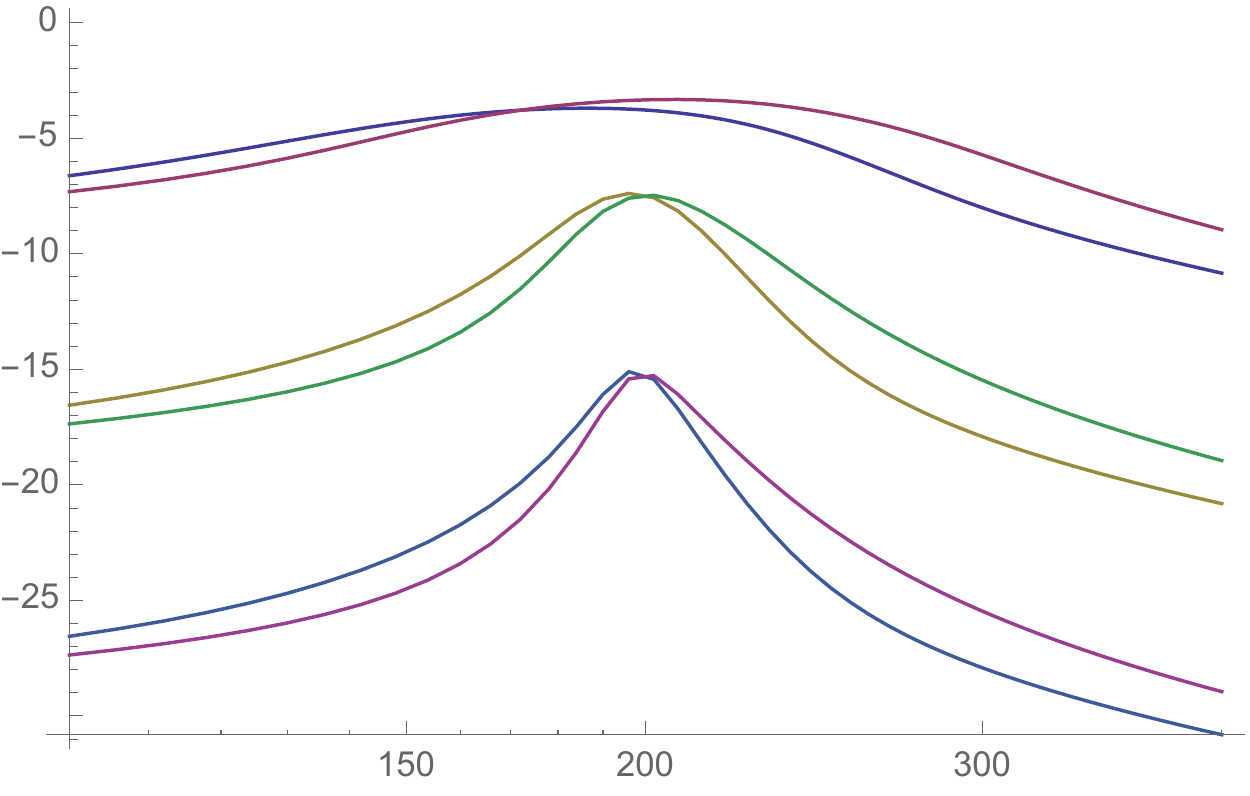}
  \caption{Influence of a stronger coupling term in Subthreshold $\mu=-0.1$ and $g=0.24$}
  \label{fig:C4}
\end{figure}

Figure \ref{fig:C4}, shows the resonance when the coupling term is strong enough. Both oscillators have the same resonant frequency, and they resonate in unison with the input signal, now the information can flow to the olfactory cortex. Now, following the analysis done by \cite{Gomez2016} it is desirable to find the coupling term $g=g_{21}=g_{12}$, for which the same resonant frequency for both oscillators appears.
The previous result motivates finding the critical coupling $g_c$ at which the systems couple. For this we write the system \ref{eq:sys2} as a set of 4 real equations, and calculate the Jacobian for this system J=

\begin{equation}
\tiny{
\left(
\begin{array}{cccc}
 -3 \omega_1 x_1^2-y_1^2 \omega_1+\mu  \omega_1 & -2 x_1 y_1 \omega_1-\omega_1 & \frac{g \omega_1}{2} & 0 \\
 \omega_1-2 x_1 y_1 \omega_1 & -\omega_1 x_1^2-3 y_1^2 \omega_1+\mu  \omega_1 & 0 & \frac{g \omega_1}{2} \\
 \frac{g \omega_2}{2} & 0 & -3 \omega_2 x_2^2-y_2^2 \omega_2+\mu  \omega_2 & -2 x_2 y_2 \omega_2-\omega_2 \\
 0 & \frac{g \omega_2}{2} & \omega_2-2 x_2 y_2 \omega_2 & -\omega_2 x_2^2-3 y_2^2 \omega_2+\mu  \omega_2 \\
\end{array}
\right)}
\end{equation}

Evaluated at the origin the Jacobian is:

\begin{equation}
J_{00}=\left(
\begin{array}{cccc}
 \mu  \omega_1 & -\omega_1 & \frac{g \omega_1}{2} & 0 \\
 \omega_1 & \mu  \omega_1 & 0 & \frac{g \omega_1}{2} \\
 \frac{g \omega_2}{2} & 0 & \mu  \omega_2 & -\omega_2 \\
 0 & \frac{g \omega_2}{2} & \omega_2 & \mu  \omega_2 \\
\end{array}
\right)
\end{equation} 

Here we can find when the eigenvalues real part is zero, to see where they cross the imaginary axis, point where the bifurcation occurs, in order to determine the critical coupling $g_c$ for the system. The four eigenvalues for the jacobian $J_{00}$ are:

\begin{align*}
 \lambda_{1,2}=\frac{1}{2} \Re\left(\mu  (\omega_1+\omega_2)\pm\sqrt{g^2 \omega_1 \omega_2+(\mu +i)^2 (\omega_1-\omega_2)^2}\right) \\
 \lambda_{3,4}=\frac{1}{2} \Re\left(\mu  (\omega_1+\omega_2)\pm\sqrt{g^2 \omega_1 \omega_2+(\mu -i)^2 (\omega_1-\omega_2)^2}\right)  
\end{align*}
Making the first two eigenvalues zero, we get the implicit equation:
\begin{equation}
\mu  (\omega_1+\omega_2)= \Re\left(\sqrt{g_c^2 \omega_1 \omega_2+(\mu +i)^2 (\omega_1-\omega_2)^2}\right)
\end{equation}
To obtain the critical value of the coupling $g_c$. This equation is correct, it is worth highlighting that the paper \citep{Gomez2016} from which this analysis is guided, has an error in this equation (equation 4 in the paper), and it does not give the desired results. However, the one presented here has been corrected.
\begin{figure}[H]
\centering
\includegraphics[width=0.5\textwidth]{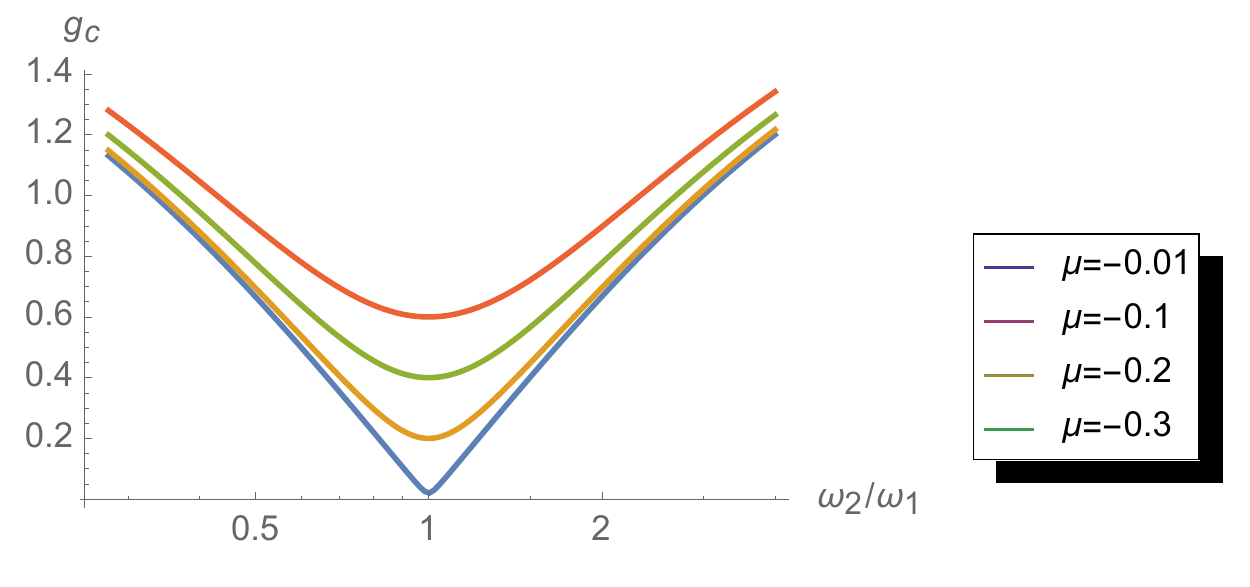}
\caption{$g_c$ as function of $\mu$ and the ratio of characteristic frequencies of the oscillators.}
\label{f:gc}
\end{figure}

In figure \ref{f:gc} the critical coupling is evaluated for different characteristic frequencies of the oscillators, and different $\mu$. It is clear, that the closer the characteristic frequencies of the system, the smaller the coupling needs to be for both systems to resonate at the same frequency, this happens for all $\mu$ in figure \ref{f:gc}.\\

\begin{figure}[H]
\centering
\includegraphics[width=0.5\textwidth]{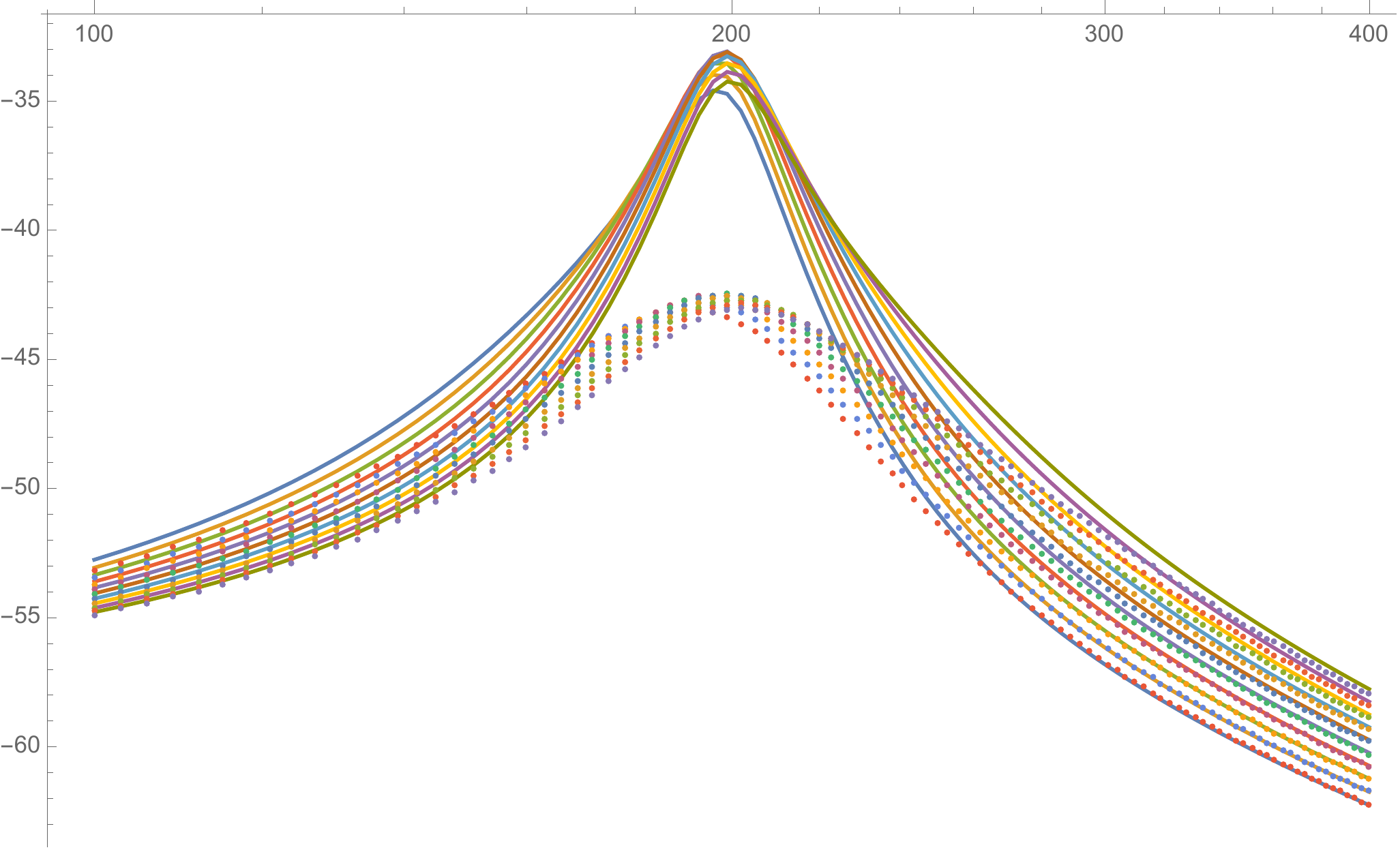}
\caption{10 oscillators with different characteristic frequency, for $\mu=-0.2$ and $\mu=-0.3$.}
\label{f:10o}
\end{figure}

We can extend this idea to systems with more elements, and different characteristic frequencies, to see if all systems oscillate at the same frequency when coupled. It can be done for 10 oscillators, in this  case for the frequencies 173, 179, 185, 191, 197, 203, 209, 215, 221, and 227 Hz. Figure \ref{f:10o} shows the coupling of the 10 oscillators for different $\mu$ values. If they were not coupled, each would resonate at their own frequency.

\subsubsection{Subcritical System, adding a 5th order term}
\begin{figure}[H]
\centering
\includegraphics[width=0.4\textwidth]{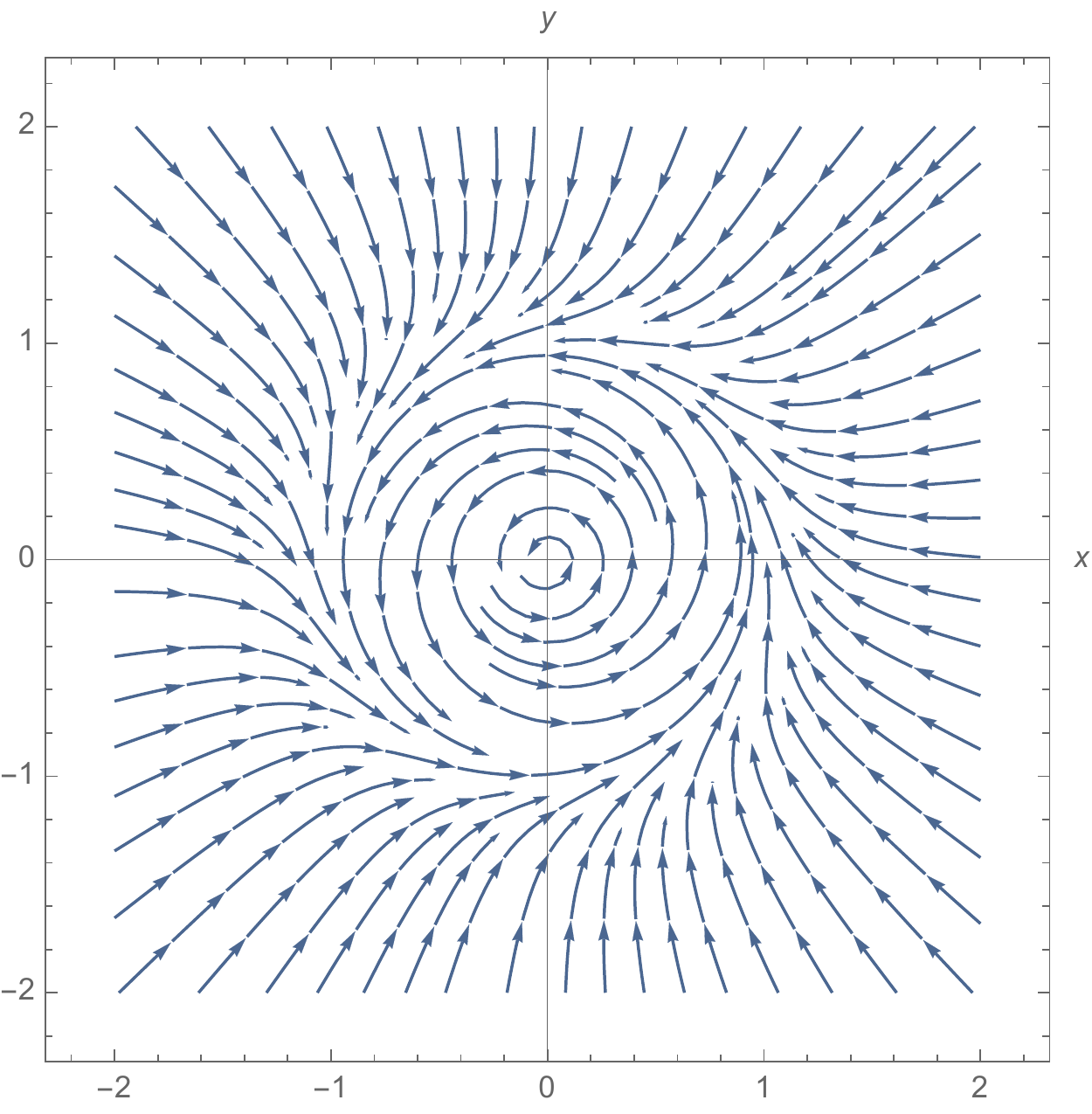}
  \caption{Behavior of the subcritical system $\mu=-0.1$ and $\omega=1$}
  \label{fig:subc1}
\end{figure}

\begin{figure}[H]
\centering
\includegraphics[width=0.4\textwidth]{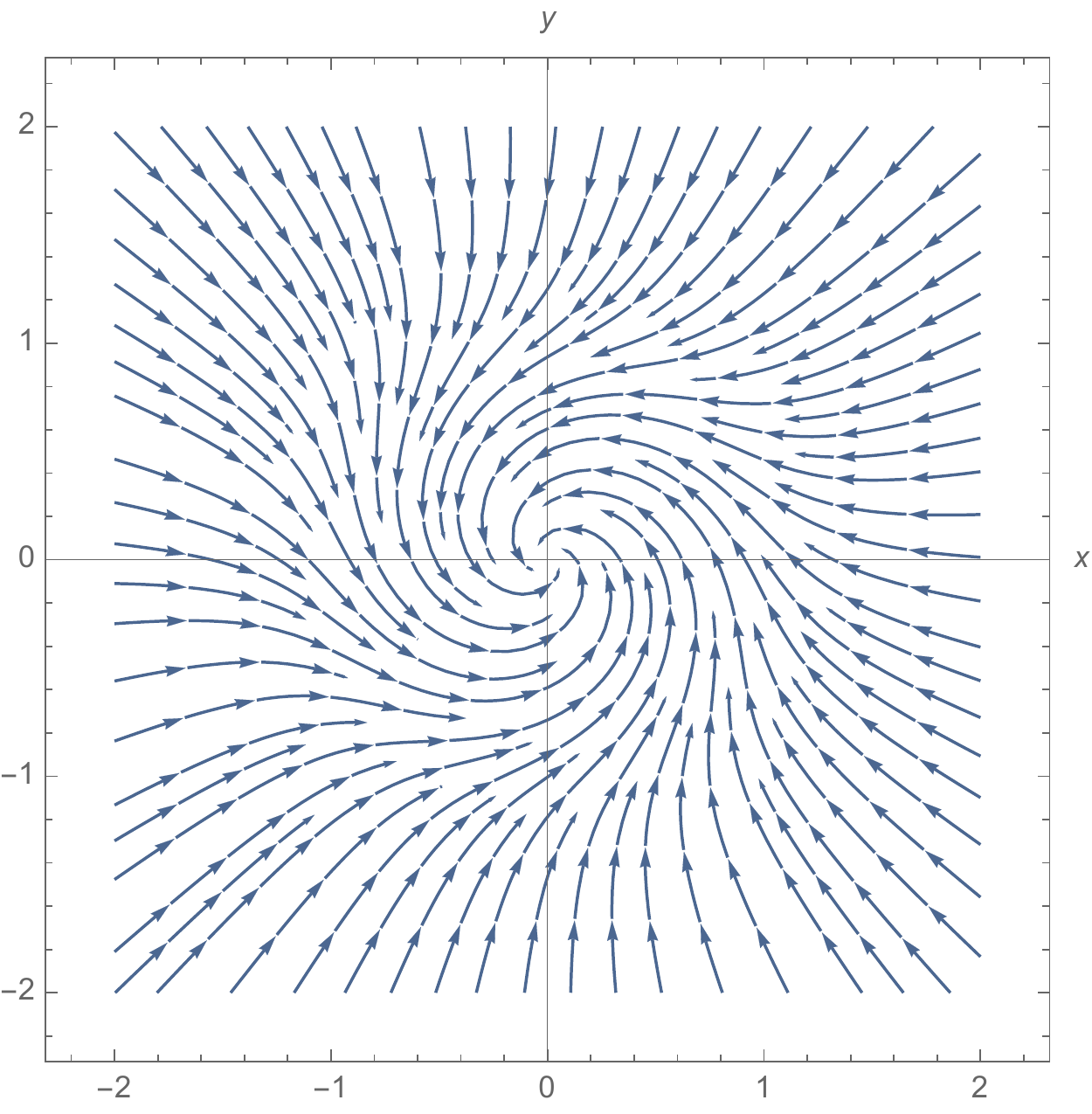}
  \caption{Behavior of the subcritical system $\mu=-0.7$ and $\omega=1$}
  \label{fig:subc2}
\end{figure}

Now let me introduce a candidate for modelling the behavior of the pyriform. When we add a 5th order term to the equation we get, the following equation:
\begin{equation}
\dot{z}=(\mu+i)\omega z+\omega |z|^2z-\omega |z|^4z
\end{equation}
Which can be separated into its real an imaginary parts getting:
\begin{align}
\dot{z}&=\dot{x}+i\dot{y} \\
\dot{x}&=-\omega x^5-2 x^3 y^2 \omega +x^3 \omega +x\mu   \omega -x y^4 \omega +x y^2 \omega -y \omega \\
\dot{y}&=-\omega  y^5 -2  y^3 x^2 \omega  +y^3\omega +y\mu  \omega    -yx^4 \omega  +yx^2 \omega  +x \omega 
\end{align}
We plot its $\dot{x}$ and $\dot{y}$ derivatives as function of $x$ and $y$, in figure \ref{fig:subc1} and \ref{fig:subc2}.

Because of the 5th order term, this system is more sensitive to the nonlinear parameter $\mu$. In figure \ref{fig:subc1} we see an stationary state for $\mu=-0.1$, this cycle is the one desirable for the behavior of the olfactory bulb, it is having the same behavior as in fig.\ref{fig:sfig2}, the above threshold behavior, but here we have the presence of cycles in the subthreshold regime and for $\mu=-0.7$ we see a fixed point at the origin. \\ \\
This capability of creating a stable orbit even at higher susceptibility to initial conditions, is a desirable behavior for modelling the pyriform or olfactory cortex given that the input odor mixture is small, variable and noise but the olfactory mechanisms manage to recognize it and creates a stable memory environment through close orbits to store them, this is a characteristic present only in the pyriform and not in the bulb, this is why the subcritical model is for the pyriform and not for the bulb. Memory is a function of the olfactory cortex, while \\

\begin{figure}[H]
\centering
\includegraphics[width=0.5\textwidth]{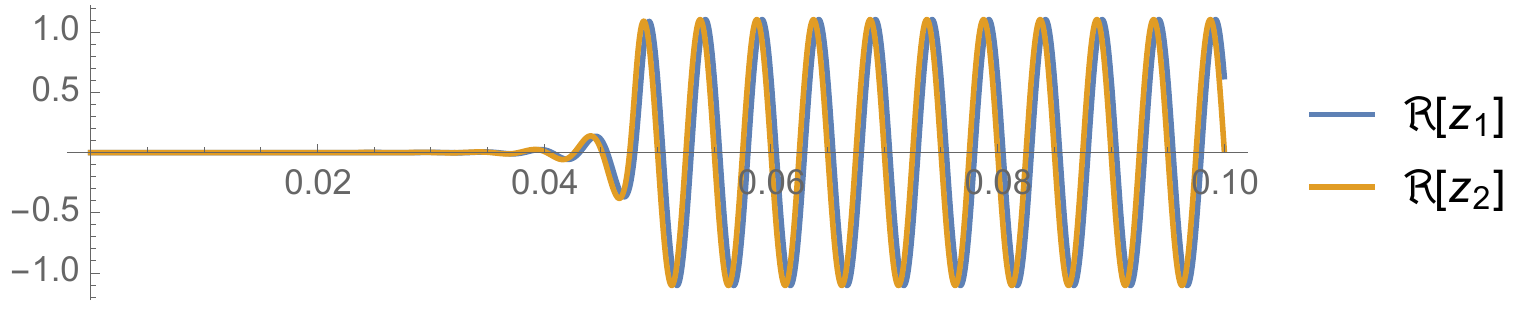}
\caption{Two coupled subcritical systems.}
\label{f:sbc}
\end{figure}

As was already mentioned in related work, the research done by \citep{Yao1990} and \citep{Li1989} suggested that the coupled system modelling the behavior of the olfactory cortex, should be able to oscillate spontaneously, that is why in the following analysis if we couple two subthreshold and subcritical hopf systems with different characteristic frequency we have: 
\begin{align}
\dot{z}_1&=\omega_1\left[(\mu+i) z_1+|z_1|^2z_1-|z_1|^4z_1+\frac{g_{21}}{2}z_2\right]\\[1ex]
\dot{z}_2&=\omega_2\left[(\mu+i) z_2+|z_2|^2z_2-|z_2|^4z_2+\frac{g_{12}}{2}z_1\right]
\label{eq:sys5ord}
\end{align}
In this case because our system oscillates spontaneously, we don't need a forcing term, but rather a very small initial condition, that should move the systems from the unstable origin. In figure \ref{f:sbc}, we see the behavior of the real parts of both systems, for $\mu_1=\mu_2=-0.1$, $\omega_1=180\cdot 2 \pi$ and $\omega_2=225 \cdot 2\pi$.

\section{Results}
\subsection{Four Subcritical coupled systems in a Network}
We now explore the behavior of 4 subcritical systems coupled on a network, each with the following equation:
\begin{equation}
\dot{z_j}=(\mu+i)\omega_j z_j+\omega_j |z_j|^2z_j-\omega_j |z_j|^4z_j+\sum_{k\neq j}^N\frac{g_{j,k}}{N}z_k
\end{equation}
Where the term $g_{j,k}$ is the graph matrix that defines the connection of the systems, $\omega_j$ is the characteristic frequency of each systems and $\mu=-0.1$ is the same for all systems. We chose different frequencies for the systems, $\omega_1=180\cdot 2\pi$, $\omega_2=220\cdot 2\pi$, $\omega_3=260\cdot 2\pi$ ans $\omega_4=300\cdot 2\pi$.\\

The network topology is bidirectional, meaning that the systems see each other with the same magnitude (weight synapse) reflecting the negative feedback that is present in the bulb and in the olfactory cortex:
\begin{figure}[H]
\centering
\includegraphics[width=0.4\textwidth]{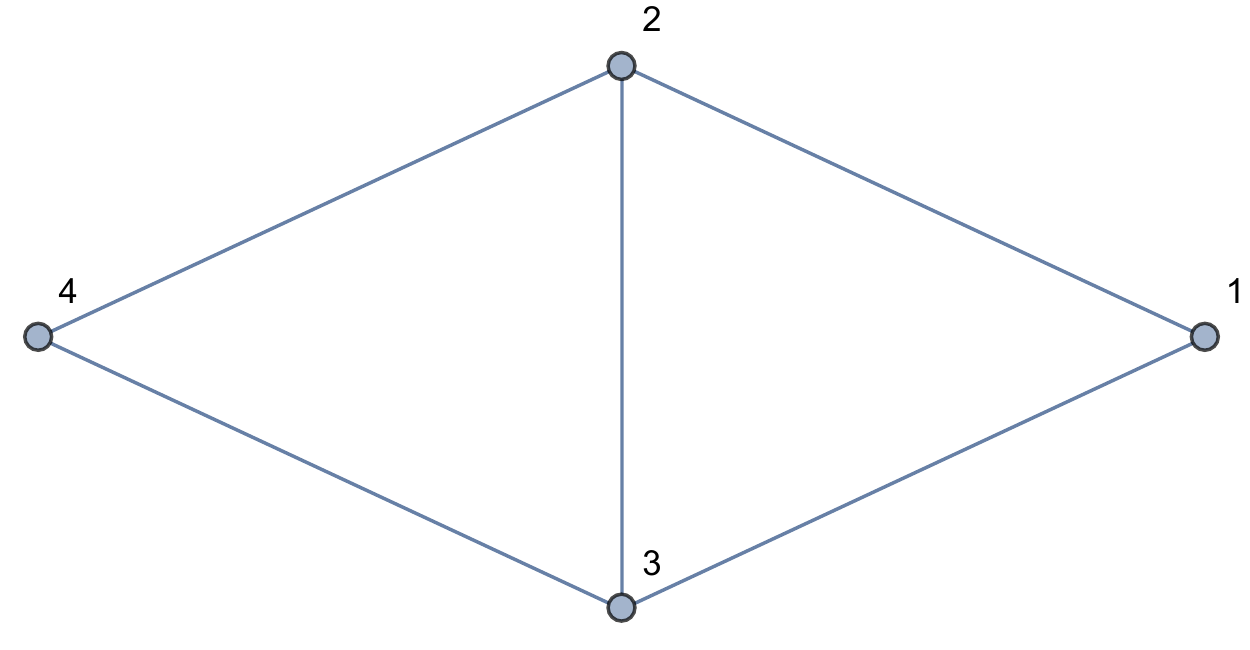}
\caption{Network topology to couple the 4 systems.}
\label{f:n4}
\end{figure}

\begin{figure*}[ht]
\centering
\includegraphics[width=\textwidth]{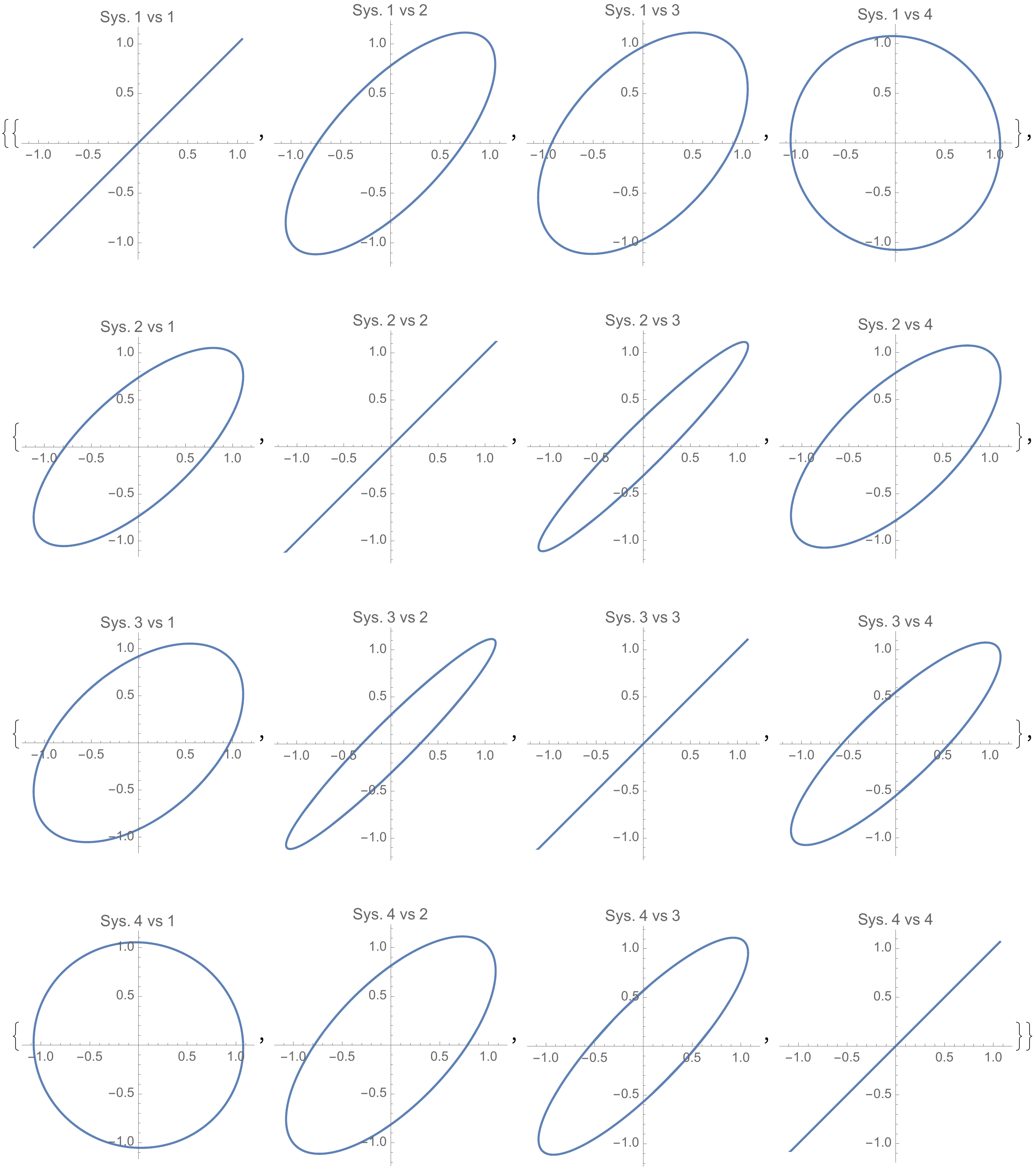}
\caption{\footnotesize Response of the real part of each system compared with the others. This is the long term response of the systems, once transient effects have dissipated.\label{f:n4c}}
\end{figure*}

The compared response of the systems can be seen in figure \ref{f:n4c}. All systems oscillate at the same frequency, however, they exhibit a phase difference, for example the phase difference between 1 and 2 is moderate, between 1 and 3 is higher, and between 1 and 4 they oscillate in opposite directions.

\begin{figure}[H]
\centering
\includegraphics[width=0.4\textwidth]{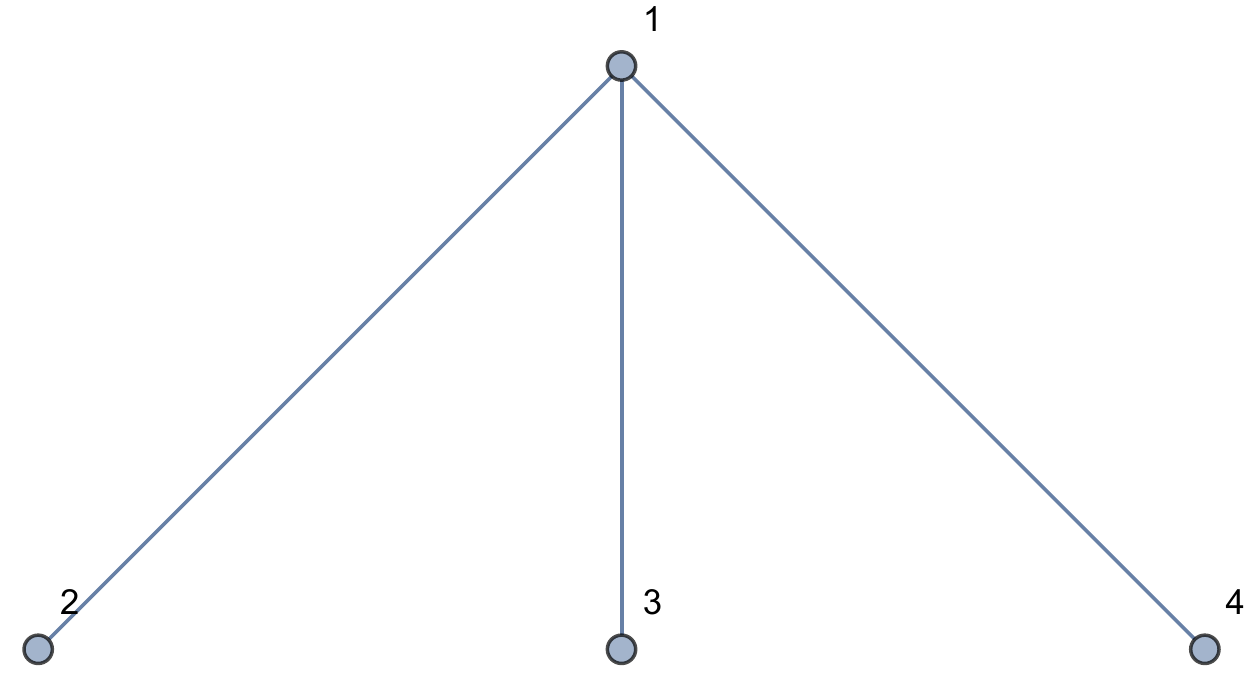}
\caption{Another network topology to couple the 4 systems.}
\label{f:n4v2}
\end{figure}

We now study the same systems but with a different network topology that can be seen in figure \ref{f:n4v2}.

\begin{figure*}[ht]
\centering
\includegraphics[width=\textwidth]{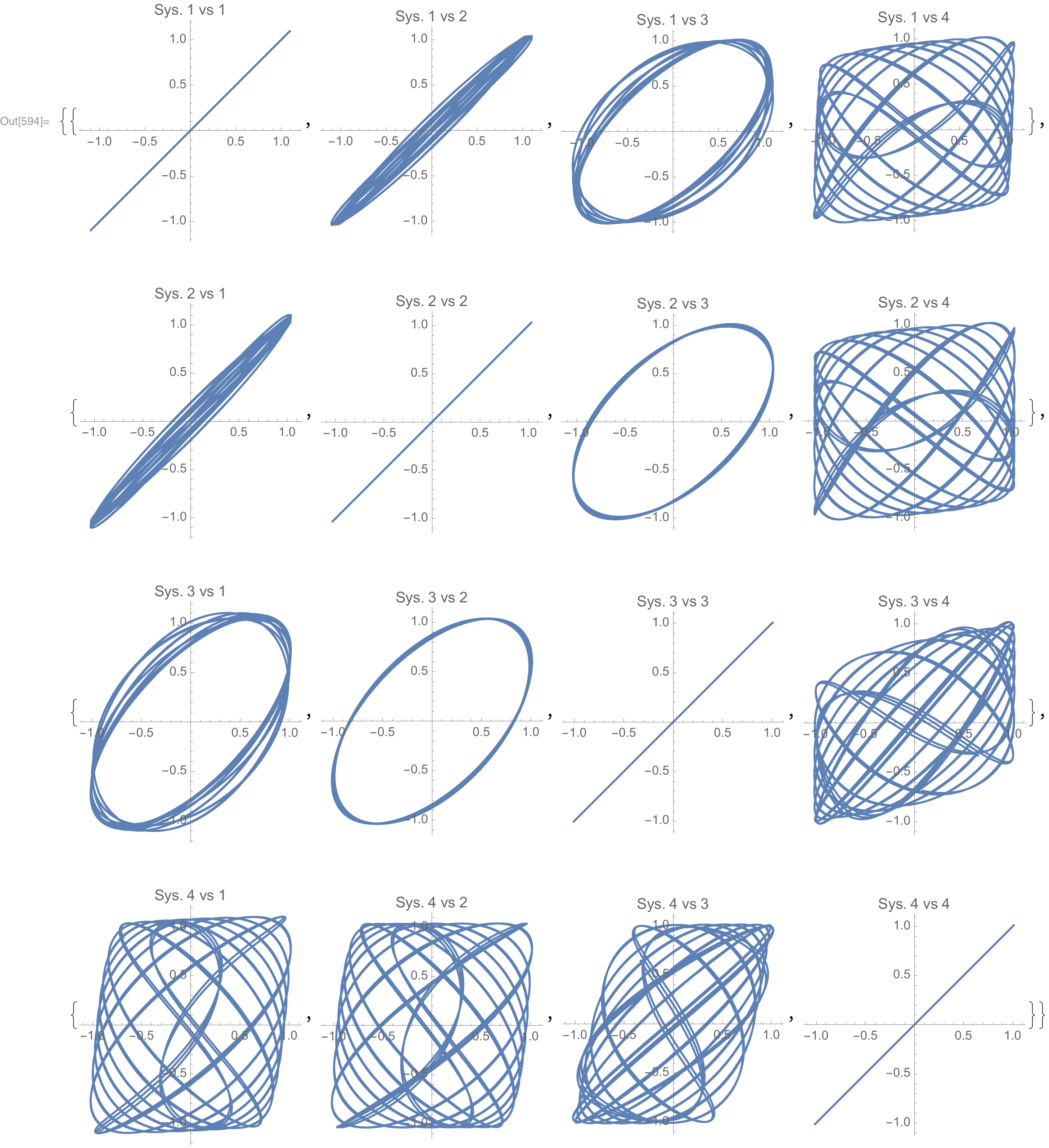}
\caption{\footnotesize Response of the real part of each system compared with the others. This is the long term response of the systems, once transient effects have dissipated\label{f:n4cv2}}
\end{figure*}

In this network topology, the only coupling that occurs happens through the first system. Therefore it is expected that the systems won't couple as easily as in the first case. In fact, this is the case the numerical solution had to  evolve for 200 times more time than in the first case. The solution to this network coupling can be seen in figure \ref{f:n4cv2}. The coupling in this case is not as strong as before, the first three systems somehow couple, however the fourth oscillates at a different frequency producing a Lissajous like behavior with the other three systems.\\

This results suggest that network topology have a strong effect on how its constituent systems couple. This reflects the importance of connectivity distribution for memory purposes in the brain, what can be translated as plasticity. in the network comparison, we kept all the conditions identical but the change in connectivity created a completely different state space. Given the memory function of the olfactory cortex, this result agrees with importance of synaptic distribution.

\section{Discussion}
This report proposes the subthreshold hopf bifurcation as a good candidate for modeling the behavior of the bulb. On the other hand, the subthreshold-subcritical hopf bifurcation is proposed as a model for the olfactory cortex. The analysis and comparison of two network topology for the subcritical model of the olfactory cortex, reflect the importance of the synaptic connectivity for memory functions.\\ \\
There is more work to be done that is proposed here (I have the results) but it is necessary to have more words in order to present them. Those are results when the subthreshold and subcritical systems are connected and when the inhibitory effect is taken into account.   

\bibliographystyle{elsarticle-harv}
\bibliography{Bib}







\end{document}